\documentclass[twoside,twocolumn,english,aps,prl]{revtex4}
\usepackage[T1]{fontenc}
\usepackage[latin1]{inputenc}
\usepackage{babel}
\usepackage{graphics}

\makeatletter
\providecommand{\LyX}{L\kern-.1667em\lower.25em\hbox{Y}\kern-.125emX\@}
\makeatother

\begin{document}

\title{Spin-entangled currents created by a triple quantum dot}

\author{Daniel S. Saraga and Daniel Loss}

\address{Department of Physics and Astronomy, University of Basel, Klingelbergstrasse
82, CH-4056 Basel, Switzerland}

\begin{abstract}
We propose a simple setup of three coupled quantum dots in the Coulomb
blockade regime as a source for spatially separated currents of spin-entangled
electrons. The entanglement originates from the singlet ground state
of a quantum dot with an even number of electrons. To preserve the
entanglement of the electron pair during its extraction to the drain
leads, the electrons are transported through secondary dots. This
prevents one-electron transport by energy mismatch, while joint transport
is resonantly enhanced by conservation of the total two-electron energy.
\\
PACS numbers : 73.63.-b, 85.35.Be, 3.65.Ud {\footnotesize \vspace{-5mm}}{\footnotesize \par}
\end{abstract}


\maketitle
The creation of entangled particles is a crucial problem as entanglement
is a prerequisite for quantum computation and communication \cite{Ben00}.
While manipulations of entangled photons demonstrating various quantum
information processing schemes have been very successful \cite{Tit01},
similar achievements are still missing for massive particles such
as electrons. Hence there has been a number of theoretical proposals
for a solid-state \emph{entangler} - a device creating two entangled
particles and allowing their separation and extraction into two distinct
channels for further processing. 

Recent proposals involved the extraction of entangled Cooper pairs
of a superconductor in contact with quantum dots \cite{Rec01},  
normal or ferromagnetic conductors 
\cite{Les01,Mel01}, and carbon
nanotubes \cite{Rec02,Ben02}. In another
scheme, the entanglement arises from interference effects in a quantum
dot in the cotunneling regime and requires special nondegenerate leads
of narrow energy width \cite{Oli02}. A generic entangler based on
interferometry and which-way detection was proposed in Ref. \cite{Bos02}.
In this article, we propose an entangler based on a triple quantum
dot setup. The entanglement originates from the singlet state of a
pair of electrons in one quantum dot, while its transport relies on
energy filtering by secondary dots. Our proposal is based on existing
technology \cite{triple} and on realistic parameter values as typically
found in transport experiments with quantum dots.

\paragraph{Setup. }

\begin{figure}[!tb]
{\centering \resizebox*{1\columnwidth}{!}{\includegraphics{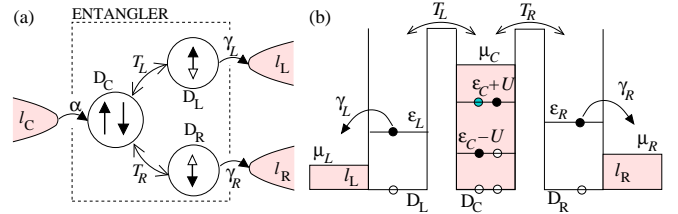}} {\footnotesize \vspace{-5mm}}\footnotesize \par}
\caption{{\footnotesize \label{figset}(a) Setup of the triple quantum dot
entangler. Three leads \protect\( l_{i}\protect \) \protect\( (i=C,L,R)\protect \)
at chemical potential \protect\( \mu _{i}\protect \) are coupled
to three quantum dots \protect\( D_{i}\protect \) in the Coulomb
blockade regime. Each dot contains an even number of electrons, and
can only accept 0, 1 (\protect\( D_{L}\protect \) and \protect\( D_{R}\protect \))
and 2 (\protect\( D_{C}\protect \)) excess electrons. A spin-singlet
is formed in \protect\( D_{C}\protect \) when two electrons tunnel
incoherently, each with a rate \protect\( \alpha \protect \), from
the source lead \protect\( l_{C}\protect \) into \protect\( D_{C}\protect \). Each of the
electrons can subsequently tunnel coherently to \protect\( D_{L}\protect \)
and \protect\( D_{R}\protect \) with tunneling amplitudes \protect\( T_{L}\protect \)
and \protect\( T_{R}\protect \). Finally, the electrons tunnel out
(with rates \protect\( \gamma _{L}\protect \) and \protect\( \gamma _{R}\protect \))
to the drain leads \protect\( l_{L}\protect \) and \protect\( l_{R}\protect \),
creating two currents of entangled electrons. (b) Energy level diagram,
for} \emph{\footnotesize each} {\footnotesize electron. Non-entangled
currents, which arise from one-electron transport, are suppressed
by the energy differences \protect\( \epsilon _{L,R}-\epsilon _{C}\pm U\protect \). 
The joint transport of both electrons is favored by conservation
of the total two-electron energy: \protect\( \epsilon _{L}+\epsilon _{R}\simeq 2\epsilon _{C}\protect \)
(resonance condition).\vspace{-5mm}}}
\end{figure}

Fig. \ref{figset} describes the proposed entangler. It
is composed of three coupled lateral quantum dots (\( D_{C},D_{L} \) and
\( D_{R} \)) in the Coulomb blockade regime, each of them coupled
to a Fermi liquid lead \( l_{C} \), \( l_{L} \) and \( l_{R} \).
When two excess electrons are present in \( D_{C} \), we can assume
\cite{singlet} that their ground state is the spin-singlet state,
which is the (anti)symmetric superposition of their (spin) wavefunctions.
The aim of the entangler is to extract the singlet from \( D_{C} \),
by transporting one electron into the neighboring dot \( D_{L} \)
and the other one into \( D_{R} \), and finally transport them into
the drain leads \( l_{L} \) and \( l_{R} \) without loss of entanglement.
This creates two currents of pairwise spin-entangled electrons that
are spatially separated.

Applying two bias voltages \( \mu _{C}-\mu _{L} \) and \( \mu _{C}-\mu _{R} \)
allows the transport of electrons from the source lead \( l_{C} \)
to both drain leads \( l_{L} \) and \( l_{R} \), via the three quantum
dots \( D_{i},i=C,L,R \). To preserve the entanglement of the electrons
until they are in the drain leads, one must avoid the individual transport
of one electron, as this would allow the arrival of a new electron
in \( D_{C} \) which could destroy the existing entanglement by forming
a new singlet with the remaining electron. To suppress one-electron
transport, we arrange the dots so that there is a large difference
between the energy levels of \( D_{L} \) and \( D_{C} \) compensated
by the energy difference between \( D_{R} \) and \( D_{C} \). This
way the joint transport of both electrons to each neighboring dot
conserves energy and is therefore enhanced by resonance, while the
off-resonant transport of one electron is suppressed by the energy
difference.

The number of electrons participating in the transport is controlled
via Coulomb blockade \cite{Kou97}, where \( N \) excess electrons
in dot \( D_{i} \) create a large electrostatic Coulomb charging
energy \( U_{i}(N) \). The energy of the \( N^{\rm th} \) electron
is then \( E_{i}(N)=U_{i}(N)-{U_{i}(N-1)}+\epsilon _{i}(N) \), where
\( \epsilon _{i}(N) \) is the lowest single-particle energy available
for the \( N^{\rm th} \) electron. We consider the electrons to be
independent and neglect further effects such as interdot charging
energy or exchange Coulomb interaction. Assuming a shell filling of
each dot \cite{Kou97}, we disregard all but the excess electrons
in each dot. The ground state in \( D_{C} \) with 2 excess electrons
is the spin singlet \( \uparrow \downarrow -\downarrow \uparrow  \),
where both electrons have the same orbital energy \( \epsilon _{C}=\epsilon _{C}(1)=\epsilon _{C}(2) \).
One excess electron in \( D_L \) or \( D_R \) cannot form a singlet with
one of the electrons already present as these are already all paired
up in singlets. We assume the gate voltages of each dot so 
that \( U_{j}(0)=U_{j}(1),j=L,R \)
and \( U_{C}(0)=U_{C}(2)=0 \), which gives a negative charging energy
for one electron: \( U_{C}(1)=-U \). We define the zero energy as
the total energy of the three empty dots.

The energy levels in the dots and the chemical potentials
\( \mu _{i} \) in the leads are assumed to be tuned such 
that only zero or one excess electron
in \( D_{j},j=L,R \), and zero, one, or two electrons in \( D_{C} \) are allowed.
It is crucial that only the ground states of the electronic levels
in the dots participate in the transport. In particular, the triplet
states in \( D_{C} \) should not be accessible to incoming electrons.
To avoid resonance with excited levels, the energy level spacings
in the dots must be larger than the Coulomb charging energies: \( \Delta \epsilon _{i}>U_{i} \).
Excited states with energy \( E_{i}^{*} \) could participate in the
transport through cotunneling events \cite{Kou97}, where one describes
the transport of one electron from \( l_{C} \) to \( l_{j} \) via
intermediate {}``virtual states{}'' in the dots by second-order
processes. We shall neglect such events, as they are suppressed by
factors of the order of \( \alpha /(E_{i}^{*}-\mu _{i})\simeq \alpha /(U_{i}-\mu _{i})\ll 1 \). 

We need low temperatures \( T \) so that thermal fluctuations cannot
allow three electrons in \( D_{C} \) or populate excited levels,
which could also create a current in the reverse direction (\( l_{j}\to D_{j} \),
\( D_{C}\to l_{C} \)). Taking \( k_{\rm B}T\ll |\mu _{i}-E_{i}(0,1,2,3)| \),
we can neglect temperature effects and set \( T=0 \) for simplicity.
The incoherent tunneling rates are given by \( \alpha =2\pi |t_{C}|^{2}\nu _{C} \)
and \( \gamma _{j}=2\pi |t_{j}|^{2}\nu _{j} \), where \( t_{i} \)
is the tunneling matrix element connecting \( l_{i} \) and \( D_{i} \),
and \( \nu _{i} \) is the density of states of the lead.

\begin{figure}[!tb]
{\centering \resizebox*{0.8\columnwidth}{!}{\includegraphics{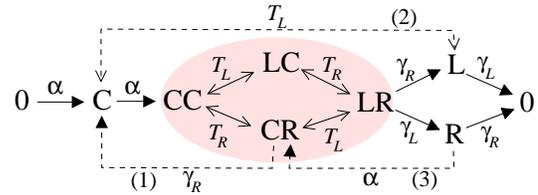}} {\footnotesize \vspace{-2mm}}\footnotesize \par}
\caption{{\footnotesize \label{figtrans} 
The entangler states \( \{0,C,CC,LC,CR,LR,L,R\} \)
and their transitions. 
Double arrows indicate coherent tunneling (oscillations) of one electron
between two dots with overlap matrix element \protect\( T_{L}\protect \)
and \protect\( T_{R}\protect \) (in the shaded area). Single arrows
indicate incoherent tunneling from (\protect\( \alpha )\protect \)
or to (\protect\( \gamma _{L}\protect \) and \protect\( \gamma _{R}\protect \))
the leads. The dashed lines indicate the three types of transitions
that must be avoided to ensure the joint transport of the singlet
pair \protect\( CC\protect \) to the leads (see text; for clarity
we do not show the transitions obtained by replacing \protect\( L\protect \)
by \protect\( R\protect \).)\vspace{-3mm}}}
\end{figure}

The quantum states of the entangler are given by combining the different
numbers of electrons allowed in each dot. \( 0 \) describes the situation
where all the dots are empty; \( L \),
\( R \) or \( C \) corresponds to one electron in \( D_{L} \), \( D_{R} \)
or \( D_{C} \), while \( CC \) denotes the singlet state created
by two electrons in \( D_{C} \). Thus, the 8 states, shown in Fig. \ref{figtrans}
with their transitions, for the basis set \(\mathcal{B}= \{0,C,CC,LC,CR,LR,L,R\}  \). This
description in terms of the individual levels of each isolated dot requires that the
tunneling matrix elements \( T_{L} \) and \( T_{R} \) (considered to be real)
connecting the dots are small and do not mix the levels in different
dots, i.e. \( T_{L},T_{R}\ll \Delta \epsilon _{i} \).

The coherent evolution is described by a Hamiltonian matrix \( H_{k,k'},k,k'\in \mathcal{B} \).
The diagonal elements contain the energies \( E_{0}=0,E_{C}=\epsilon _{C}-U,E_{CC}=2\epsilon _{C},E_{LC}=\epsilon _{L}+\epsilon _{C}-U,E_{CR}=\epsilon _{R}+\epsilon _{C}-U,E_{LR}=\epsilon _{L}+\epsilon _{R},E_{L}=\epsilon _{L},E_{R}=\epsilon _{R} \),
while the off-diagonal elements describe the coherent oscillations
of one electron between \( D_{C} \) and \( D_{L} \) or \( D_{R} \):
\( H_{C,L}=H_{CR,LR}=T_{L},H_{C,R}=H_{LC,LR}=T_{R},H_{CC,LC}=T_{L}\sqrt{2} \)
and \( H_{CC,CR}=T_{R}\sqrt{2} \). The \( \sqrt{2} \) factor comes
from the identical orbital states in \( CC \). 

We describe the incoherent  transport as sequential tunneling 
(lowest order in  \( \alpha ,\gamma _{L},\gamma _{R} \)) in terms of the
master equation for the (reduced)
density matrix
\( \rho  \) of the entangler. The diagonal elements, \( \rho _{k},k\in \mathcal{B} \),
are the occupation probabilities of the state \( k \), with normalization \( \sum _{k}\rho
_{k}=1 \). The off-diagonal elements \( \rho _{k,k'} \) contain the coherent
superposition of \( k \) and \( k' \). We write the master equation
\cite{Blu96} as \( \mathrm{d}\rho /\mathrm{dt}=-i[H,\rho ]-M \), where
\( M \) describes the incoherent transport from/to a lead connecting
a state \( k \) to another state \( k' \) with a rate 
\( W(k',k)\in \{\alpha ,\gamma _{L},\gamma _{R}\} \).
For the diagonal elements, this results in the population equation:
\( M_{k}=-\rho _{k}\sum _{k'}W(k',k)+\sum _{k'}W(k,k')\rho _{k'} \),
while the off-diagonal elements are damped by the incoherent transitions
out of each of the two corresponding states: \( M_{k,k'}=-\frac{1}{2}\rho _{k,k'}\sum _{k''\neq k,k'}W(k'',k)+W(k'',k') \).
Out of the 64 elements of \( \rho  \), only \( 24 \) are coupled
to the relevant diagonal elements. Arranging them in a real vector
\( \overrightarrow{V} \), one can rewrite the master equation as
a homogeneous first-order differential equation given by a matrix
\( \mathcal{A} \): \( \mathrm{d}\overrightarrow{V}/\mathrm{dt}=\mathcal{A}\overrightarrow{V} \).
Its stationary solution is the eigenvector corresponding to the zero
eigenvalue of \( \mathcal{A} \). It can be found symbolically by
a mathematical software (MAPLE), and defines the stationary populations
\( P_{k}=\rho ^{0}_{k} \) of the different states \( k \) of the
entangler. 

\paragraph*{Results.}

Our aim is to show that the transport of the singlet state through
the entangler is robust and dominates over the transport of uncorrelated
electrons. We define the stationary entangled current (see the central
sequence in Fig. \ref{figtrans}) as \( I_{E}=e\gamma _{L}(P_{LR}+P_{L})+e\gamma _{R}(P_{LR}+P_{R}) \).
The destruction of the entanglement of the two electrons can occur
in three ways; see Fig. \ref{figtrans} (the following discussion
also refers to the cases with \( L \) replaced by \(  R \)). 

(1) An electron tunnels out from \( D_{R} \) to \( l_{R} \) while
the second electron is still in \( D_{C} \). This creates a current
\( e\gamma _{R}P_{CR} \) which might contain no entanglement, as
the remaining electron can form a new singlet with a new electron
coming from \( l_{C} \). This current is suppressed by the energy
mismatch \( \epsilon _{R}-\epsilon _{C}-U \) between the states \( CC \)
and \( CR \). The entangled current \( I_{E} \) is dominant if the
\emph{entangler quality} \( Q=\min \{P_{LR}/P_{CR},P_{LR}/P_{LC}\} \)
satisfies \( Q\gg 1 \).

(2) An electron tunnels from \( D_{C} \) to \( D_{L} \) and then
out to \( l_{L} \) before a second electron has tunneled into \( D_{C} \)
and formed a singlet. This channel is suppressed by the energy difference
\( \epsilon _{L}-\epsilon _{C}+U \) between the states \( C \) and
\( L \). As the stationary solution contains no information on the
history of the electrons, we compare currents obtained in the two
following cases. \emph{(i)} We switch off the undesired channel by
setting \( T_{L}=T_{R}=0 \) between \( C \), \( L \) and \( R \),
while keeping the tunneling between \( CC,CR,LC \), and \( LR \).
This defines the stationary populations \( P_{k} \). \emph{(ii)}
We keep the undesired channel, while switching off the tunneling involving
\( CC \) and \( LR \). This defines the populations \( \tilde{P}_{k} \),
and creates a current \( e\gamma _{L}(\tilde{P}_{L}+\tilde{P}_{LC})+e\gamma _{R}(\tilde{P}_{R}+\tilde{P}_{CR}) \)
containing no entanglement. Defining a second entangler quality \emph{}by
\( \widetilde{Q}=\min \{P_{L}/\tilde{P}_{L},P_{LR}/\tilde{P}_{LC},P_{R}/\tilde{P}_{R},P_{LR}/\tilde{P}_{CR}\} \),
the condition \( \widetilde{Q}\gg 1 \) corresponds to the suppression
of this one-electron channel.

(3) After the joint transport of the two electrons into state \( LR \)
and the tunneling of one electron into \( l_{L} \), a new electron
tunnels into \( D_{C} \) before the remaining electron in \( D_{R} \)
has tunneled out to \( l_{R} \). The new electron can then form a
new singlet with the remaining electron, therefore destroying the
entanglement that existed with the electron which has moved to the
lead \( l_{L} \). To suppress this channel, we need \( \alpha \ll \gamma _{L},\gamma _{R} \).
For simplicity, we set \( \alpha =0 \) for transitions from \( L \)
and \( R \) when calculating the probabilities \( P_{k} \) (but
we keep \( \alpha  \) non-zero for \( \tilde{P}_{k} \)).

The exact analytical expressions for \( P_{k} \) are extremely lenghty
and cannot be written in a compact form. We introduce \( \delta \epsilon _{j}=\epsilon _{j}-\epsilon _{C} \),
which explicitly removes the dependence on \( \epsilon _{C} \) as
only energy differences enter the time evolution of the density
matrix \( \rho (t) \). Secondly, we consider a symmetric setup with
\( \gamma =\gamma _{L}=\gamma _{R} \) and \( T_{0}=T_{L}=T_{R} \).
Then, the symmetry between \( LR \), \( L \), and \( R \) yields
\( P_{L}=P_{R}=P_{LR} \) (see Fig. \ref{figtrans}). Thirdly, we
consider the case \( \epsilon _{R}=\epsilon _{C} \). The resonance
condition for the joint transport \( E_{CC}=2\epsilon _{C}\simeq E_{LR}=\epsilon _{L}+\epsilon _{R} \)
translates into \( \delta \epsilon _{L}\simeq 0 \), and the energy
differences relevant for the suppression of one-electron transport
are given by \( U \) and \( U\pm \delta \epsilon _{L} \). From a
qualitative analysis involving first and second-order perturbation
calculations (\emph{\`a la} Fermi Golden Rule), we find that the conditions
for a dominant joint transport read \( \alpha \ll \gamma \ll T_{0}\ll U,|U\pm \delta \epsilon _{L}| \)
\cite{notefgr}. Hence we expand both the numerator and denominator
of the expressions for \( P_{k} \) in the lowest non-trivial order
in \( \alpha ,\gamma ,T_{0} \). We distinguish two cases:
\\
{[}I{]} non-resonant (\( |\delta \epsilon _{L}|>T_{0} \)):\begin{eqnarray}
P_{L}=P_{R}=P_{LR} & \simeq  & \frac{2T_{0}^{4}(2U-\delta \epsilon _{L})^{2}}{\delta \epsilon_{L}^{2}U^{2}(U-\delta \epsilon _{L})^{2}}\, ,\label{eqa1} \\
P_{CR} & \simeq  & \frac{2T_{0}^{2}}{U^{2}}\, \, ,\, \, P_{LC}\simeq \frac{2T_{0}^{2}}{(U-\delta \epsilon _{L})^{2}}\, ,\label{eqa2} \\
\tilde{P}_{R} & \simeq  & \frac{T_{0}^{2}}{U^{2}}\, \, ,\, \, \tilde{P}_{L}\simeq \frac{T_{0}^{2}}{(U+\delta \epsilon _{L})^{2}}\, ,\label{eqa3} 
\end{eqnarray}
 {[}II{]} resonant (\( \delta \epsilon _{L}=0 \)):\begin{eqnarray}
P_{L}=P_{R}=P_{LR} & \simeq  & \frac{8\alpha T_{0}^{4}}{\alpha \gamma ^{2}U^{2}+4T_{0}^{4}(8\gamma +9\alpha )}\, ,\label{eqb1} \\
P_{CR}=P_{LC} & \simeq  & \frac{2\alpha T_{0}^{2}(\gamma ^{2}U^{2}+40T_{0}^{4})}{U^{2}\left[ \alpha \gamma ^{2}U^{2}+4T_{0}^{4}(8\gamma +9\alpha )\right] }\, ,\label{eqb2} \\
\tilde{P}_{R}=\tilde{P}_{L} & \simeq  & \frac{\alpha \gamma T_{0}^{2}}{\alpha \gamma U^{2}+2T_{0}^{2}(\gamma +\alpha )^{2}}\, ,\label{eqb3} 
\end{eqnarray}
and \( \tilde{P}_{CR}=\tilde{P}_{R}\alpha /\gamma  \), \( \tilde{P}_{LC}=\tilde{P}_{L}\alpha /\gamma  \). 

\begin{figure}[!b]
{\centering {\footnotesize \vspace{-0mm}}\resizebox*{1\columnwidth}{!}{\rotatebox{-90}{\includegraphics{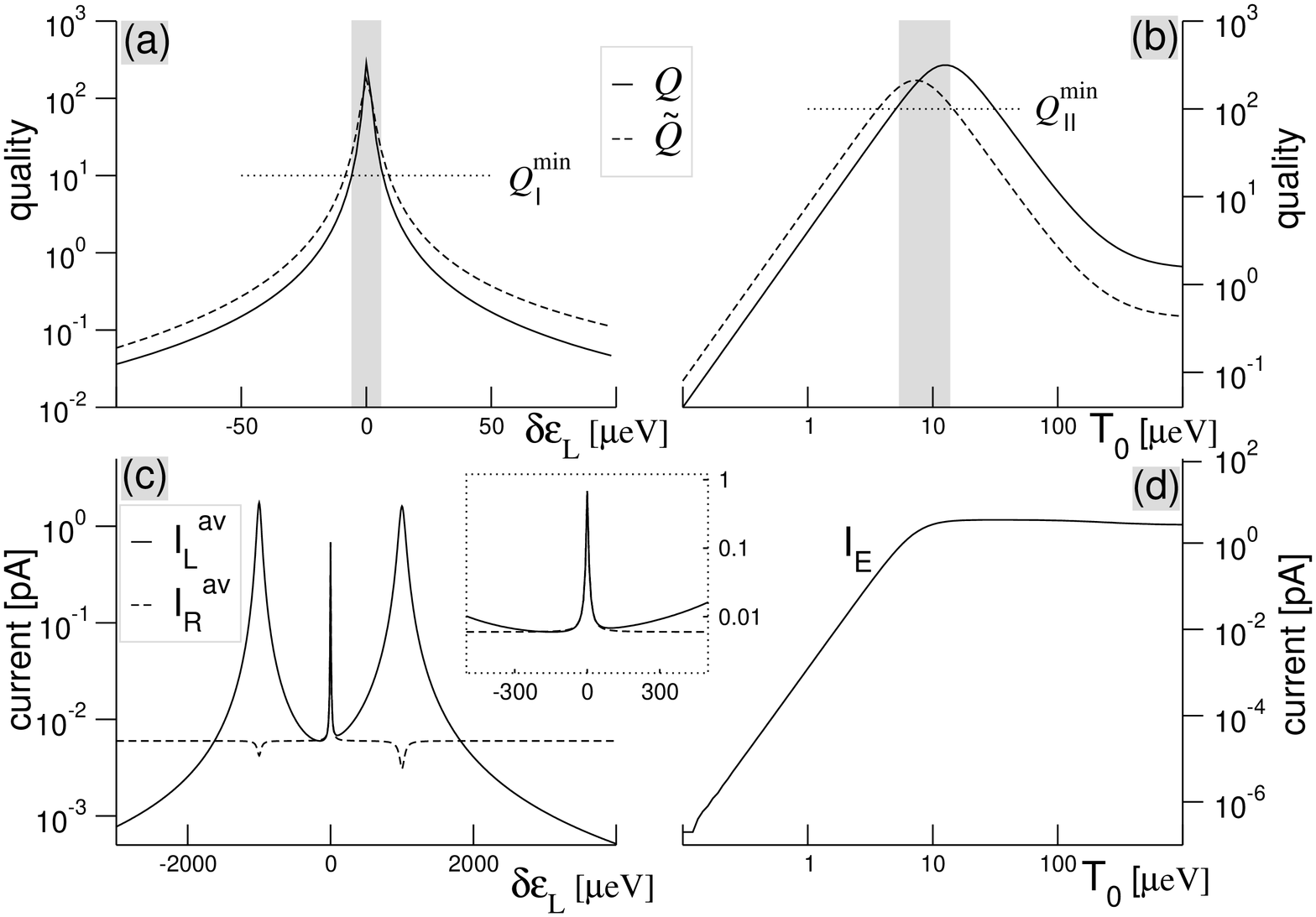}}} {\footnotesize \vspace{-5mm}}\footnotesize \par}
\caption{\label{figres}{\small Quality of the entangler and output current,
with the parameters \protect\( \alpha =0.1,\gamma =1,T_{0}=10,U=1000\protect \)
in \protect\( \mu \mathrm{eV}\protect \). (a) Quality \protect\( Q\protect \)
and \protect\( \widetilde{Q}\protect \), around the resonance at \protect\( \delta \epsilon _{L}=0\protect \)
where the entangled current dominates. The width of the resonance
defined by \protect\( Q,\widetilde{Q}>Q_{\mathrm{I}}^{\mathrm{min}}=10\protect \)
is \protect\( |\delta \epsilon _{L}|<  \overline{\delta \epsilon _{L}}  =2T_{0}/\sqrt{Q_{\mathrm{I}}^{\mathrm{min}}}\simeq 6\, \mu \mathrm{eV}\protect \)
(shown in gray). (b) \protect\( Q\protect \) and \protect\( \widetilde{Q}\protect \)
as a function of the tunneling matrix element \protect\( T_{0}\protect \)
at resonance (\protect\( \delta \epsilon _{L}=0\protect \)). In gray,
the region where the quality of the entangler is \protect\( Q,\widetilde{Q}>Q_{\mathrm{II}}^{\mathrm{min}}=100\protect \).
(c) Average current in the left (\protect\( I^{\mathrm{av}}_{L}\protect \))
and in the right (\protect\( I_{R}^{\mathrm{av}}\protect \)) leads,
which takes into account both entangled and non-entangled currents.
A symmetric current (\protect\( I^{\mathrm{av}}_{L}=I^{\mathrm{av}}_{R}\protect \))
is the signature of the resonance and therefore of the desired regime
where the entangled currents dominate (see the inset with a larger
scale). (d)} \emph{\small }{\small At resonance, the entangled current
\protect\( I_{E}\protect \) saturates to \protect\( \simeq \alpha =4\, \mathrm{pA}\protect \)
when \protect\( T_{0}\gg (U^{2}\gamma \alpha /32)^{1/4}\simeq 5\, \mu \mathrm{eV}\protect \).
\vspace{-5mm}}}
\end{figure}

These two cases are sufficient for an analytical discussion of the
entangler, as the approximate expressions (\ref{eqa1})-(\ref{eqb3})
reproduce accurately the \emph{exact} results presented in Fig. \ref{figres}.
As shown in Fig. \ref{figres}(a), the entangler qualities \( Q \)
and \( \widetilde{Q} \) reach a maximum around \( \delta \epsilon _{L}=0 \),
which is due to a resonance in the coherent oscillations between \( CC \)
and \( LR \). 
 We define now the quantities \( Q_{\mathrm{I}} \) and
\( \widetilde{Q}_{\mathrm{I}} \) as approximations of the qualities
\( Q \) and \( \widetilde{Q} \) obtained with Eqs. (\ref{eqa1})-(\ref{eqa3}).
\( Q_{\mathrm{I}} \) and \( \widetilde{Q}_{\mathrm{I}} \),
which grow as \( (\delta \epsilon _{L})^{-2} \), give a correct estimate
of the width \( \overline{\delta \epsilon _{L}} \) of the resonance
around \( \delta \epsilon _{L}=0 \). Introducing the condition \( Q_{\mathrm{I}},\widetilde{Q}_{\mathrm{I}}>Q_{\mathrm{I}}^{\mathrm{min}} \),
we get \( |\delta \epsilon _{L}|<\overline{\delta \epsilon _{L}}=2T_{0}/\sqrt{Q_{\mathrm{I}}^{\mathrm{min}}} \).
Note that such proportionality of the width to the tunneling matrix
element  is also found in the Rabi formula for a two-level system.
Similarly, we define \( Q_{\mathrm{II}} \) and \( \widetilde{Q}_{\mathrm{II}} \)
with Eqs. (\ref{eqb1})-(\ref{eqb3}); these approximate ratios accurately
reproduce the height of the resonance peak at \( \delta \epsilon _{L}=0 \).
Introducing the condition \( Q_{\mathrm{II}},\widetilde{Q}_{\mathrm{II}}>Q_{\mathrm{II}}^{\mathrm{min}} \),
we find \( \gamma \sqrt{Q_{\mathrm{II}}^{\mathrm{min}}/8}<T_{0}<U\sqrt{4\alpha /\gamma Q_{\mathrm{II}}^{\mathrm{min}}} \);
see Fig. \ref{figres}(b). 

Quantities which are experimentally accessible are the currents in
the left and right leads, \( I_{L}=e\gamma _{L}(P_{LC}+P_{LR}+P_{L}) \),
and \( I_{R}=e\gamma _{R}(P_{CR}+P_{LR}+P_{R}) \). Far from resonance
(\( \delta \epsilon _{L}>T_{0} \)), one cannot neglect the currents
\( \tilde{I}_{L} \) and \( \tilde{I}_{R} \) coming from the single-electron
transitions described by the probabilities \( \tilde{P}_{k} \). Hence
we consider in Fig. \ref{figres}(c) the average of the currents in
situations \emph{(i)} and \emph{(ii)}: \( I_{L}^{\mathrm{av}}=(I_{L}+\tilde{I}_{L})/2 \)
and \( I_{R}^{\mathrm{av}}=(I_{R}+\tilde{I}_{R})/2 \). Away from
the two-electron resonance the current is asymmetric: \( I_{L}^{\mathrm{av}}\neq I_{R}^{\mathrm{av}} \).
The large peaks where \( I_{L}^{\mathrm{av}}\gg I_{R}^{\mathrm{av}} \)
are due to one-electron resonances: between \( CC \) and \( LC \)
when \( \delta \epsilon _{L}=U \) (right peak), and between \( C \)
and \( L \) when \( \delta \epsilon _{L}=-U \) (left peak). At the
\( \delta \epsilon _{L}=0 \) resonance the current is symmetric as
the electrons are transported together and simultaneously from the
central dot to the leads {[}see inset in Fig. \ref{figres}(c){]}.
Hence by varying \( \epsilon _{L} \) via the gate voltage of \( D_{L} \)
until \( I_{L}=I_{R} \), one can locate the resonance, and therefore
the regime where the entangler is most efficient \cite{asym}. Finally,
Eq. (\ref{eqb1}) gives \( P_{LR}\to \alpha /4\gamma  \) for \( T^{4}_{0}\gg U^{2}\gamma \alpha /32 \),
which yields \( I_{E}\to I_{\mathrm{max}}=\alpha  \), as illustrated
in Fig. \ref{figres}(d). Note that in order to be able to reach 
\( I_{\mathrm{max}} \) within the window
 \( \gamma \sqrt{Q_{\mathrm{II}}^{\mathrm{min}}/8}<T_{0}<U\sqrt{\alpha /4\gamma Q_{\mathrm{II}}^{\mathrm{min}}} \),
one also needs the condition \( U>Q_{\mathrm{II}}^{\mathrm{min}}\gamma \sqrt{\gamma /2\alpha } \).

\paragraph*{Discussion.}

Setting the qualities of the entangler to \( Q_{\mathrm{II}}^{\mathrm{min}}=100 \),
\( Q_{\mathrm{I}}^{\mathrm{min}}=10 \) and \( \gamma =10\alpha  \),
we need, approximately, \( 35\alpha <T_{0}<U/60 \), and \( U>2200\alpha  \).
The first condition is easily met as \( \alpha  \) and \( T_{0} \)
can be varied via the voltages defining the barriers \cite{Exp}.
For the second condition, present-day experiments manipulate currents
typically around \( 2\, \mathrm{pA} \) \cite{Exp}, yielding \( \alpha >0.1\, \mu \mathrm{eV} \)\( \Rightarrow  \)\( U>0.3\, \mathrm{meV} \).
To get a finite width \( \delta \epsilon _{L}\simeq 6\mu \mathrm{eV} \),
we take \( U\simeq 1\, \mathrm{meV} \), which is within reported
values \cite{Exp,Kou97}. To obtain an even better quality (\( Q_{\mathrm{II}}^{\mathrm{min}}=1000,\gamma =100\alpha  \)),
one needs to increase the ratio \( U/\alpha  \). One possible issue
is the single-particle energy spacing \( \Delta \epsilon _{i} \),
which is usually smaller than or equal to the charging energy. However,
as \( \Delta \epsilon _{i}\propto 1/L^{2} \) in a box of size \( L \),
while \( U\propto 1/L \), one should be able to reach \( \Delta \epsilon _{i} \gg U \)
by decreasing the dot size. Alternatively, one can use vertical quantum
dots, which have large energy level spacings \cite{Kou97}, or use
one carbon nanotube with two bendings (which defines three regions
behaving like quantum dots). 

A current \( I_{\mathrm{max}}\simeq 10\, \mathrm{pA} \) corresponds
to the delivery of an entangled pair every \( t_{\mathrm{p}}\simeq 1/\alpha \simeq 16\, \mathrm{ns} \),
which is below reported spin decoherence time of \( 100\, \mathrm{ns} \)
\cite{Kik97}. The average separation between two entangled electrons
is approximately \( t_{\mathrm{e}}\simeq 1/U\simeq 0.4\, \mathrm{ps}\ll t_{\mathrm{p}} \),
with a maximal separation of \( t_{\mathrm{m}}\simeq 1/\gamma \simeq 0.5\, \mathrm{ns} \).
This would allow noise measurements using a beam splitter, where an
enhancement in the two-terminal noise is a signature of singlet states
compared to the noise of non-correlated electrons \cite{Bur00}. Note
that such an experiment requires electrons with the same orbital energy,
which can be achieved if \( \epsilon _{L}=\epsilon _{R}=\epsilon _{C} \).
One could also carry out a measurement of Bell's inequalities by measuring
the spin of the electron in each lead with the help of a spin filter
based on a quantum dot \cite{Rec00}. 

\begin{acknowledgments}
{\footnotesize \vspace{-0mm}}We thank H.A. Engel, P. Recher, C. Egues,
and G. Burkard for useful discussions. Financial support from NCCR
{}``Nanoscale Science{}'', the Swiss NSF, DARPA, and ARO is gratefully
acknowledged.{\footnotesize \vspace{-5mm}} 
\end{acknowledgments}


\begin{thebibliography}{10}
\bibitem{notefgr}Although the requirement \( \gamma \ll T_{0} \) is not consistent
with the conditions of validity of a perturbation theory in \( T_{0} \),
such a calculation clarifies the physical picture of the entangler by
showing explicitly the suppression due to the energy mismatch.
\bibitem{asym}The entangled current has a small asymmetry (\( \sim 5\%) \) at resonance
if \( \gamma _{L}\neq \gamma _{R},T_{L}\neq T_{R} \).
\end{thebibliography}
\end{document}